\PassOptionsToPackage{table}{xcolor}
\documentclass[sigconf]{acmart}
\AtBeginDocument{%
  }

\renewcommand\footnotetextcopyrightpermission[1]{} 

\usepackage{amsmath,amsfonts}
\usepackage{algorithmic}
\usepackage{algorithm}
\usepackage{array}
\usepackage{textcomp}
\usepackage{stfloats}
\usepackage{url}
\usepackage{verbatim}
\usepackage{graphicx}

\usepackage{cleveref}
\crefname{figure}{Fig.}{Figs.}
\crefname{table}{Table}{Tables}

\usepackage{amsmath}
\usepackage{booktabs}
\usepackage{multirow}
\usepackage[table]{xcolor}

\newcommand{\eg}{\textit{e.g.}}

\begin{document}
\title{Unified Steganography via Implicit Neural Representation}


\author{
Qi Song$^{1}$, \
Ziyuan Luo$^{1}$, \
Xiufeng Hunag$^{1}$, \
Sheng Li$^{2}$, \
Renjie Wan$^{1}$\hspace{-0.15cm} \
\vspace{0.1cm}
}
\authornote{Corresponding author}

\author{
$^{1}$Department of Computer Science, Hong Kong Baptist University \\
$^{2}$School of Computer Science, Fudan University \\
\vspace{0.05cm}
\texttt{\{qisong,ziyuanluo,xiufenghuang\}@life.hkbu.edu.hk} \\
\texttt{lisheng@fudan.edu.cn}, \ \texttt{renjiewan@hkbu.edu.hk}
}

\renewcommand{\shortauthors}{ }

\begin{abstract}
 Digital steganography is the practice of concealing for encrypted data transmission. Typically, steganography methods embed secret data into cover data to create stega data that incorporates hidden secret data. However, steganography techniques often require designing specific frameworks for each data type, which restricts their generalizability. In this paper, we present U-INR, a novel method for steganography via Implicit Neural Representation (INR). Rather than using the specific framework for each data format, we directly use the neurons of the INR network to represent the secret data and cover data across different data types. To achieve this idea, a private key is shared between the data sender and receivers. Such a private key can be used to determine the position of secret data in INR networks. To effectively leverage this key, we further introduce a key-based selection strategy that can be used to determine the position within the INRs for data storage.  Comprehensive experiments across multiple data types, including images, videos, audio, and SDF and NeRF, demonstrate the generalizability and effectiveness of U-INR, emphasizing its potential for improving data security and privacy in various applications.
\end{abstract}


\maketitle
\pagestyle{plain} 

\section{Introduction}
\label{sec:intro}

Digital steganography aims to hide information within digital data, such as images, audio, or video, to achieve encrypted transmission of information. Typically, steganography methods~\cite{zhu2018hidden_r77,jing2021hinet,zhang2019steganogan_rr73,li2023steganerf,mstafa2020new_video_stega} embed \textbf{secret data} into \textbf{cover data} to create \textbf{stega data} that incorporates  hidden secret information. Receivers can extract previously hidden secret data from transmitted stega data.

However, these approaches are often tailored to a specific data format, such as images~\cite{jing2021hinet,li2024purified_pusnet}, audio~\cite{djebbar2012comparative_audio_stega}, or video~\cite{mou2023video_stega,weng2019high,jia2022rivie_video} media. For each data type, current solutions~\cite{zhu2018hidden_r77,jing2021hinet,mou2023video_stega,cvejic2002increasing_lsb_audio} have to design specific encoders and extractors to process the data, which lacks the flexibility to be universally applied across different data types. For example, the frameworks~\cite{li2024purified_pusnet,jing2021hinet} designed for image hiding cannot be used for audio or video steganography. This limitation poses a significant challenge for users who want to hide data across various media types without being restricted to specific data formats. Besides, existing methods heavily rely on external extractors to extract hidden information, which introduces critical security vulnerabilities. Attackers could exploit these extractors to corrupt~\cite{zhu2021destroying,geetha2021steganogram} or expose~\cite{liu2022whendeep,yang2023provably} the secret data. Such inherent insecurity fundamentally undermines the reliability of conventional steganography in practical deployments, where robustness against malicious attacks is paramount.

\begin{figure}[t]
\centering
\includegraphics[width=0.95\columnwidth]{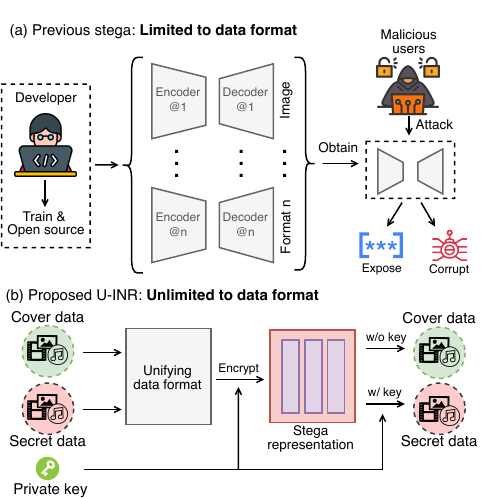}
\vspace{-0.35cm}
\caption{Illustration of our scenario. (a) Previous steganography approaches require designing specific frameworks for different data formats. Besides, malicious users could exploit the steganography encoder/decoder to expose or corrupt the secret data. (b) Our \textbf{U-INR} can work on various data formats like image, video and others. Besides, U-INR bypasses the need for external encoder/decoder architectures that could expose
attack surfaces, ensuring exclusive access to key holders.} 
\vspace{-0.5cm}
\label{fig1}
\end{figure}

We envision a new scenario where secret data can be embedded into cover data without data-type constraints. As demonstrated in \cref{fig1}, in our new scenario, the data hiding and extraction do not rely on format-specific encoders and extractors, preventing the limitations of specific data formats in previous steganography methods. The data senders and receivers use a private key that defines how secret data is embedded and extracted, eliminating the reliance and the risks of attacks on the external components.
This scenario bypasses the need for external encoder/decoder architectures that could expose attack surfaces, ensuring exclusive access to key holders and preserving confidentiality and security.

We propose to overcome the challenges posed by various data formats through representing different types of data in a unified manner, thereby eliminating the need to manage different data formats. Our approach employs Implicit Neural Representations (INRs)~\cite{sitzmann2020siren}, which inherently satisfy this requirement by providing a unified framework for encoding diverse multimedia data (\eg, images, audio, videos and others).  Besides, prior works~\cite{frankle2020pruning,han2015learning} have established that neural networks contain significant redundancy, with certain neurons being removable without degrading model performance. As INR naturally owns the network-based representation capability, it motivates us to use separate neurons in the INR to store the cover data and the hiding data, which enables seamless information embedding while preserving the perceptual quality of the cover media.

INR's distinctive capability has inspired serval works for data hiding, yet existing methods~\cite{liu2023_stegaINR,dong2024implicit_stega,chen2024nerf_stega} still remain limited to single modalities, overlooking INRs' across-modality representation capability~\cite{sitzmann2020siren,mildenhall2020nerf}. Recent
INRSteg~\cite{yang2023_INRSteg} suggests combining INRs for data steganography across different data modality. However, this approach explicitly modifies the default network structure of the INR, making it noticeable to attentive attackers. Besides, since the user must know the details of modifications, such a strategy is not conducive to the recipient's retrieval. The above weakness underscores a necessity for a novel method that is less detectable by attentive attackers and a more appropriate way for receivers to retrieve the hidden data.

To address the above problems, we propose a novel strategy in which receivers and senders share a consensus. This consensus establishes a pre-defined agreement between sender and receiver that identifies specific neurons for encoding cover data versus those allocated for secret data embedding. Then, the sender and receivers can use such a consensus for data hiding and extraction, respectively. In our design, the identified neurons within the INR are utilized to represent the secret data, while the remaining neurons are optimized to encode the cover data. We refer to this optimized INR with the cover and secrete data as a \textbf{stega representation}. During the extraction, this method minimizes detectability by vigilant attackers and simplifies the retrieval process for receivers, ensuring a more efficient and secure steganography.

We introduce an \textbf{implicit consensus mechanism}, which can effectively identify the positions of the parameters storing the secret data.  In our design, the private shared key helps identify the positions used for hiding secret data. We achieve this by using this key to initialize the weights of the INR and then sorting and selecting the weights based on their values. The data sender and receiver can use this key to achieve data hiding and retrieval. Unauthorized users without the correct keys cannot obtain the secret data from the stega representation. We call this strategy \textbf{U-INR}, which allows users to conceal data within the parameters of INR and extract the hidden data using a designated key. By leveraging INR’s inherent flexibility and adaptability, our approach facilitates the seamless integration of hidden data into the existing data structure, enabling more secure communication and data transmission. In summary, our contributions can be summarized as:
\begin{itemize} 
\item We introduce a novel steganography framework, unifying multimedia data representation for data hiding and retrieval.
\item We develop a parameter-level embedding technique that integrates secret data directly within INRs, enabling cross-modal steganography without dedicated encoders or format-specific extractors.
\item We introduce an implicit consensus mechanism to securely identify hidden data positions, enhancing data transmission safety against unauthorized access.
\end{itemize}
Extensive experiments have been conducted across various INR-based representations to demonstrate the generalizability and effectiveness of our method, resulting in an advanced improvement compared to existing steganography methods.

\section{Related work}

\subsection{Traditional steganography}
Steganography~\cite{tao2018towards,zhang2011reversible,zhang2006efficient,shi2016reversible,yang2023provably,zhang2018robust_tmm} involves concealing secret data within a carrier medium, creating a covert information container~\cite{morkel2005overview_image_stega,djebbar2012comparative_audio_stega}. In image steganography, a cover image serves as the vessel for embedding a secret image~\cite{baluja2017hiding_rr5}. Traditional approaches, such as spatial-based techniques~\cite{imaizumi2014multibit_rr24,nguyen2006multi_rr41,pan2011image_rr43,provos2003hide_rr46,jiang2024robust}, often employ strategies like Least Significant Bits (LSB), pixel value differencing (PVD)~\cite{pan2011image_rr43}, and manipulation of multiple bit-planes~\cite{nguyen2006multi_rr41} or color palettes~\cite{imaizumi2014multibit_rr24,niimi2002high_rr42,hou2018reversible}. However, these methods can introduce statistical anomalies that are detectable by steganalysis tools. To better prevent detection, adaptive strategies~\cite{pevny2010using_rr45,li2014new_rr31} have been developed. These strategies focus on making the presence of secret data more invisible by minimizing embedding distortion and optimizing data coding to maintain visual indistinguishability. Similarly, transform-based methods~\cite{chanu2012image_rr10,kadhim2019comprehensive_rr27}, including JSteg~\cite{provos2003hide_rr46} and Discrete Cosine Transform (DCT) steganography~\cite{hetzl2005graph_rr21,briassouli2005hidden_dct}, have struggled to achieve substantial payload capacities. Recent advancements in steganographic techniques, leveraging the power of deep learning~\cite{zhu2018hidden_r77,jing2021hinet,zhang2019steganogan_rr73,li2023steganerf,mstafa2020new_video_stega,mou2023video_stega}, significantly enhance the capacity and security of data concealment. These methods employ neural networks~\cite{li2024cover,li2023towards,luo2023securing,li2023steganography} to intricately analyze and subtly modify complex data attributes, resulting in more sophisticated and less detectable forms of data steganography. Baluja~\cite{baluja2017hiding_rr5} pioneers a deep-learning approach capable of embedding a full-sized image within another. GAN~\cite{shi2018ssgan_rr53} is used to create synthetic container images with probability map techniques for minimal distortion embedding~\cite{pevny2010using_rr45,tang2017automatic_rr59}. U-Net generators~\cite{yang2019embedding_rr69} integrated with adversarial frameworks have also been introduced, targeting distortion minimization\cite{tang2019cnn_rr58}. Additionally, three-player game models such as SteganoGAN~\cite{zhang2019steganogan_rr73} and HiDDeN~\cite{zhu2018hidden_r77} employ auto-encoder architectures in an adversarial manner to enhance resistance to steganalysis.  PUSNet~\cite{li2024purified_pusnet} proposes extracting a steganography network from a common network, enabling the covert transmission of the steganography network. 
However, these methods overlook the potential of using neurons directly to represent secret data for steganography. The type of data limits these methods, as the corresponding pipeline must be designed according to each data type.

\subsection{INR-based steganography}
Implicit Neural Representation (INR) utilizes neural networks to learn continuous functions for data like images and shapes~\cite{sitzmann2020siren,mildenhall2020nerf}. It has advanced generative modeling, 3D reconstruction~\cite{li2024variational}, and compression, demonstrating high-quality results from limited data~\cite{chen2019learning_r1,sitzmann2020siren,dupont2021coin_r6,niemeyer2019occupancy_r8}. INR provides a groundbreaking approach to data representation by enabling continuous functions across multiple data modalities. Studies~\cite{yang2023_INRSteg,liu2023_stegaINR, biswal2024Steganerv,dong2024implicit_stega,chen2024nerf_stega} explore the use of implicit neural representations (INR) for steganography.  Existing works \cite{luo2023copyrnerf,song2024protecting} attempt INR-based steganography, they still treat INR like traditional data formats~\cite{zhu2018hidden_r77,morkel2005overview_image_stega}, where secret data is extracted from the outputs of INR~\cite{mildenhall2020nerf,sitzmann2020siren}. StegaNeRV~\cite{biswal2024Steganerv} uses an additional extractor to retrieve hidden information from the reconstructed results, ignoring the possibility of directly using neurons to represent the hidden information. Using partial neuron weights to represent the secret data is a straightforward idea~\cite{yang2023_INRSteg,liu2023_stegaINR,dong2024implicit_stega,chen2024nerf_stega}. Methods like ~\cite{liu2023_stegaINR,dong2024implicit_stega,chen2024nerf_stega} address data steganography within a single modality, overlooking the capability of INR methods~\cite{sitzmann2020siren} to represent multiple data types. The recent INRSteg~\cite{yang2023_INRSteg} proposes combining several different INRs to achieve steganography, but this changes the original network structure of the INR, making it easily noticeable by attentive attackers. Moreover, this concatenation method is not conducive to the recipient's retrieval, as the user must know which neurons were used to store the hidden information. This highlights the need for a novel method that is more imperceptible to attentive attackers and easier for recipients to retrieve.

\begin{figure*}[t]
\centering
\includegraphics[width=1.00\linewidth]{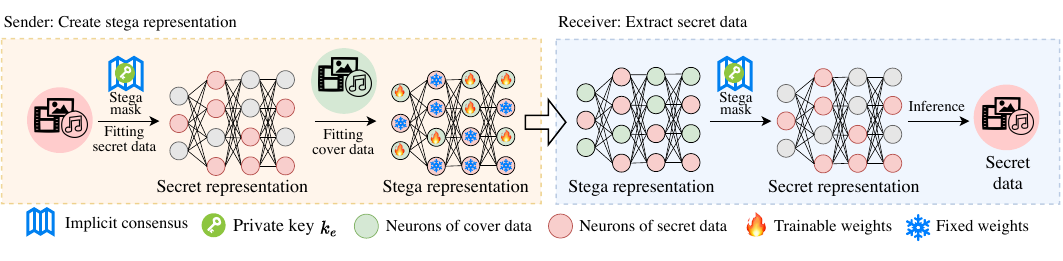} 
\vspace{-1.1cm}
\caption{ Framework of U-INR. Our architecture establishes secure synchronization between multimedia sender and receiver through an implicit consensus using a shared private key $k_e$. This key precisely maps the weight positions distinguishing secret payloads from cover data in the neural representation. Enforcing consensus-based parameter coordination through the implicit neural network's weight-sharing mechanism eliminates the need for external auxiliary modules.}
\vspace{-0.4cm}
\label{fig:framework}
\end{figure*}

\section{Problem Formulation}

\subsubsection{Traditional steganography} Traditional steganography frameworks are inherently constrained by data format dependencies, requiring specialized encoders and extractors for each media type. Given a cover data \( C \) (\eg, an image, audio, or video) and a secret data \( S \), the objective is to generate stega data \( C^* \) using an embedding function tailored for the specific data format:
\begin{equation}
C^* = E(C, S).
\end{equation}
Once the stega data \( C^* \) is received, the secret message \( S \) is extracted using a format-specific extractor:
\begin{equation}
S = D(C^*).
\end{equation}

Traditional steganography methods are fundamentally limited by their dependence on specific data formats. Developing a format-specific framework requires significant effort as each media type (\eg, images, audio, video) has unique characteristics to implement steganography. This results in a lack of flexibility, preventing the seamless application of steganography techniques across different media types. Besides, the reliance on these specialized extractors introduces critical vulnerabilities. Attackers can target these extractors to access or corrupt the hidden. This reliance on $D$ introduces critical vulnerabilities, as attackers can exploit these extractors to access or corrupt the secret data~\cite{zhu2021destroying,geetha2021steganogram,liu2022whendeep,yang2023provably}. Each tailored extractor $D$ represents a potential attack surface, increasing the risk of security breaches. Consequently, the robustness of traditional steganography methods is compromised, particularly in environments where security against malicious attacks is paramount.

\subsubsection{Our proposed scenario}
We propose a unified steganography paradigm via Implicit Neural Representations (INRs) to transform multimedia data from various modalities into a unified representation. Our approach operates directly on the INR's parameter space, eliminating the need for format-specific operations.

Consider an INR network parameterized by $\Theta$, representing the cover data \( C \). The secret message \( S \) is embedded directly within the parameters of $\Theta$:
\begin{equation}
\Theta^* = \Theta(C, S, k_e),
\end{equation}
where $k_e$ is a private key identifying the positions within the parameter space where the secret data is stored. The stega data
$C^*$ is represented by the INR network $\Theta^*$:
\begin{equation}
C^* = \Theta^*(\cdot).
\end{equation}
The secret data $S$ could be retrieved using the extraction function from the stega INR \( \Theta^* \):
\begin{equation}
S = \Theta^*(k_e).
\end{equation}
The private key $k_e$ ensures that only authorized users can access the secret message, thereby enhancing security.

Our framework takes advantage of the unique representation capability of INR, offering a unified and efficient solution for steganography across multiple media types.  As the need for external extractors is eliminated, we also address the critical vulnerability~\cite{zhu2021destroying,geetha2021steganogram,liu2022whendeep,yang2023provably} in traditional steganography. 

\section{Proposed method}
The overview framework of U-INR is depicted in \cref{fig:framework}. We utilize Implicit Neural Representation (INR) to transform multimedia data from various modalities into a unified representation. This enables a unified steganography framework applicable to diverse data types. Then, instead of depending on external modules for hiding and extraction, we introduce an implicit consensus mechanism between data senders and receivers. This mechanism supports both data hiding and extraction, mitigating risks associated with external components. 

\subsection{Unified multimedia data representation}
Traditional steganography methods are inherently constrained to multimedia data types due to their reliance on format-specific frameworks. These approaches struggle to generalize across modalities as format-specific frameworks cannot harmonize heterogeneous data. In contrast, our work leverages Implicit Neural Representations (INRs)~\cite{sitzmann2020siren,mildenhall2020nerf} to dissolve such weakness. We propose encoding diverse multimedia data into a unified neural representation, adapting to arbitrary data types like images, audio, and 3D scenes seamlessly.

INR employs continuous functions representing various data formats, such as images, videos, audio, and 3D scenes.
An INR inputs spatial or temporal coordinates and outputs the corresponding information at the coordinates, such as image pixel values or amplitude for audio. Generally, an INR $\mathbf{F}_\Theta$ can be denoted as:
\begin{equation}
\mathbf{y} = \mathbf{F}_\Theta(\mathbf{u}) ,
\label{eq:inr_inference}
\end{equation}
\noindent where $\mathbf{u}$ represents the input coordinates, which can be either spatial (for images and 3D scenes) or temporal (for audio), $\theta$ denotes the learned parameters of the neural network, and $\mathbf{y}$ is the output information such as pixel or amplitude values. This framework allows a unified approach to modeling different data types using a flexible neural network-based function $\mathbf{F}_\Theta$. For instance, in the context of image representation, an INR would receive a 2D coordinate $\mathbf{u} \in \mathbf{R}^2$ and output a pixel value $\mathbf{c} \in \mathbf{R}^3$. Similarly, for audio signals, a one-dimensional temporal coordinate $t \in \mathbf{R}$ would be mapped to an audio amplitude $\mathbf{a} \in \mathbf{R}$. INR has already demonstrated strong representational capabilities across various modalities of data.  Given these unique advantages, we propose leveraging INR to unify multimedia data format. Then, the data sender and receiver are relieved from directly handling the format of steganographic data. They only need to determine the position of the neurons for secret data, thereby alleviating the threat caused by external encoders/extractors.

\subsection{Implicit consensus mechanism}
\label{sub_method_1}
The primary motivation behind our implicit consensus stems from a critical vulnerability in traditional steganographic approaches: their dependence on external explicit extractors~\cite{li2024purified_pusnet,zhu2018hidden_r77}. These external explicit extractors present a significant security weakness—they serve as a clear attack surface that adversaries can exploit to detect, corrupt, or expose hidden information~\cite{liu2022whendeep,yang2023provably,zhu2021destroying,geetha2021steganogram}. We alleviate such vulnerability by designing an implicit data hiding and extraction mechanism. Our approach lets the sender and receiver cryptographically agree on which parameters contain hidden data using only a pre-shared private key $k_e$ without requiring any external extraction mechanism during transmission. This shift from explicit extractors to implicit consensus significantly strengthens security by removing a primary attack vector. Moreover, this approach inherently supports format-agnostic steganography, as the implicit mechanism operates on neural parameters regardless of the underlying data modality, addressing both the security and flexibility limitations that have constrained traditional steganographic methods. The whole process for making the implicit consensus is as follows.

\subsubsection{Obtaining private key} In this work, the private key replaces the external extraction mechanism during the data distribution. This private key enables both the hiding and extraction of data within our framework. Specifically, the private key determines the locations for data hiding. Prior to distribution, data senders utilize the private key to generate the stega representation. Subsequently, the sender uses a designated method to securely share the private key with the recipients. Upon receiving the steganographic data, the recipients apply the pre-shared private key to retrieve the hidden information. In our setup, we use a pre-arranged \textbf{private key $\mathbf{k_e}$}. Rather than expli
citly transmitting the secret data's position along with the cover representation, users can use pre-shared Arabic numerals as the private key $k_e$, and only receivers with the correct key can access the secret data. 


\begin{figure}[t]
\centering
\includegraphics[width=1.00\linewidth]{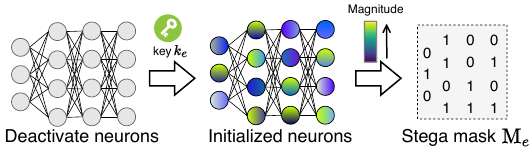} 
\vspace{-0.75cm}
\caption{Implicit Consensus. The initialized weight values of the INR are used to identify and select the weights for secret data based on the magnitudes.}
\label{fig:framework_2}
\vspace{-0.5cm}
\end{figure}

\subsubsection{Creating stega mask $\mathrm{M}_e$}

With the private key $k_e$, we can create a stega mask $\mathrm{M}_e$ by determining the position of the secret data.
The stega mask maintains a binary matrix corresponding to neuron positions, enabling selective neural modulation for secure data embedding.  The stega mask $\mathrm{M}_e$ can be used to extract the secret data as it records the position of the secret data. 
Compared to the previous methods~\cite{liu2023_stegaINR,dong2024implicit_stega,chen2024nerf_stega}, our approach allows receivers to directly obtain the location of secret data based on the pre-arranged key $\mathbf{k_e}$, thereby avoiding potential risks during data transmission. Once the stega data is received, these positions can be retrievable to ensure the receiver can regain access to the secret data. This way, the sender and receiver can reliably identify and extract the secret information embedded within the model's parameters using the shared key.

With the private key $\mathbf{k_e}$, the data sender could effectively determine the position of secret data via our proposed \textbf{implicit consensus} (\cref{fig:framework_2}). The algorithm details of implicit consensus are depicted in \cref{algoritm}. For an implicit neural network $\mathbb{N}[\mathrm{W}](\cdot)$, where $\mathbb{N}[\cdot]$ and $\mathrm{W}$ denote the architecture and weights respectively. we have the secret representation $\mathbb{N}[\mathrm{W} \odot \mathrm{M}_e](\cdot)$. Here, $\odot$ is the element-wise product, and $\mathrm{M}_e$ is a stega mask indicating secret representation positions. We leverage initialized weights from the seed key $k_e$. This method selects significant weights by magnitude, preserving network performance. Given a stega ratio $\mathcal{S}$ and total weight count $\mathcal{N}$, the algorithm initializes weights $\mathrm{W}_e$ using $\mathbb{N}[\cdot]$ and $k_e$, ensuring reproducibility. Weights are sorted by absolute value to identify significance. A threshold $t_{\mathcal{S}}$ is set as the $p$-th largest weight, where $p = \lfloor \mathcal{S} \cdot \mathcal{N} \rfloor$ and $\lfloor \rfloor$ is floor function, selecting the top $\mathcal{S} \cdot 100\%$ weights. The stega mask $\mathrm{M}_e$ is generated by marking weights exceeding $t_{\mathcal{S}}$ as $1$, others as $0$. This binary mask highlights significant weights, encoding the secret representation effectively. 

This approach serves multiple purposes. First, it ensures that only the most impactful weights are retained, which helps maintain the model's performance while embedding the secret representation. Second, by focusing on weight magnitude, we avoid the pitfalls of selecting arbitrary weights, which could lead to suboptimal performance or even model failure. Lastly, this method provides a systematic and reproducible way to generate the stega mask, making it easier to maintain consistency across different model iterations and experiments. Overall, the implicit mechanism enhances the robustness and reliability of the U-INR framework.

\begin{algorithm}[t]
\caption{Creating stega mask $\mathrm{M}_e$ via implicit consensus}
\label{algoritm}

\begin{algorithmic}[1]
\REQUIRE Neural network architecture $\mathbb{N}[\cdot]$, seed key $k_e$, stega ratio $\mathcal{S}$, total weight count $\mathcal{N}$, initialization function $\mathcal{I}$
\ENSURE Stega mask $\mathrm{M}_e$

\STATE \textbf{Initialize Weights:}
\STATE Initialize weights $\mathrm{W}_e$ using $\mathcal{I}(\mathbb{N}[\cdot], k_e)$

\STATE \textbf{Sort Weights:}
\STATE Sort the weights $\mathrm{W}_e$ in descending order based on their absolute values.

\STATE \textbf{Determine Threshold:}
\STATE Calculate $p = \lfloor \mathcal{S} \cdot \mathcal{N} \rfloor$
\STATE Identify the threshold $t_{\mathcal{S}}$ as the value of the $p$-th largest weight in $\mathrm{W}_e$

\STATE \textbf{Generate Stega Mask:}
\FOR{each weight $w_i$ in $\mathrm{W}_e$}
    \IF{$|w_i| > t_{\mathcal{S}}$}
        \STATE Set $\mathrm{M}_e[i] = 1$
    \ELSE
        \STATE Set $\mathrm{M}_e[i] = 0$
    \ENDIF
\ENDFOR

\STATE \textbf{Output the Stega Mask:}
\RETURN $\mathrm{M}_e$

\end{algorithmic}
\end{algorithm}

\subsection{Encrypting stega representation}
\label{sub_method_2}

After obtaining the stega mask $\mathrm{M}_e$, users can construct the stega representation using secret and cover data. To obtain the secret representation $\mathbb{N}[\mathrm{W} \odot \mathrm{M}_e]$, we first optimize the representation for the secret data based on the stega mask $\mathrm{M}_e$. Next, we fix these optimized secret weights and optimize the remaining weights $\mathbb{N}[\mathrm{W} \odot \overline{\mathrm{M}}_e]$ for the secrete data to create the final stage representation. The following sections provide a detailed illustration of this process.

\subsubsection{Fitting secret data.}
Our weight selection strategy gives a choice to obtain the stega mask $\mathrm{M}_e$ based on the private key $k_e$. Given secret data (\eg, image, video, or other types), its information values $\mathbf{y}_{se}$ and corresponding coordinate $\mathbf{u}$, we can build
the secret representation by 
\begin{equation}
\hat{\mathbf{y}}_{se} = \mathbb{N}[\mathrm{W}\odot\mathrm{M}_e](\mathbf{u}),
\label{eq:secret}
\end{equation}
where $\hat{\mathbf{y}}_{se}$ is the predicted secret data, and
the secret weights determined by stage mask $\mathrm{M}_e$ are optimized via \cref{eq:optimize}.

\subsubsection{Fitting cover data.}
After optimizing the weights for the secret data, we fix those weights and update the remaining weights to store the cover data. The process can be denoted as
\begin{equation}
    \hat{\mathbf{y}}_{st} = \mathbb{N}[\mathrm{W}_\text{fix} \odot \mathrm{M}_e \cup \mathrm{W} \odot \overline{\mathrm{M}}_e](\mathbf{u}),
\label{eq:cover}
\end{equation}
where $\overline{\mathrm{M}}_e$ is a binary mask complementing $\mathrm{M}_e$, and $\mathrm{y}_{st}$ refers to the stega data at corresponding coordinate $\mathbf{u}$.

The above process can be achieved via standard INR procedure~\cite{sitzmann2020siren,mildenhall2020nerf}, which can be formulate as 
\begin{equation}
\mathcal{L} = \frac{1}{N} \sum_{i=1}^{N} (y_{gt}^{(i)} - \hat{y}^{(i)})^2 ,
\label{eq:optimize}
\end{equation}
where $N$ is the number of samples, $y_{gt}^{(i)}$ is the ground truth value for the i-th sample, and $\hat{y}^{(i)}$ is the predicted value for the i-th sample. When the optimization settles down, we can get the stega representation with both cover and secret data. The implementation details of each data modality are provided in \cref{sec:details}.

\subsection{Decrypting secret representation}
\label{sub_method_3}
After obtaining the stega representation, the ordinary users without the correct key can only obtain the whole data stored in the stega representation via standard INR inference \cref{eq:inr_inference} as:
\begin{equation}
    \mathbb{N}[\mathrm{W}](\mathbf{u}) \to \hat{\mathbf{y}}_{st},
\end{equation}
where $\hat{\mathbf{y}}_{st}$ is the output values of stega data at coordinates $\mathbf{u}$.
Ordinary users can only obtain stega data through standard inference as secret data is stored in particular INR neurons. Even if attackers suspect hidden data exists, searching the high-dimensional neuron space becomes computationally prohibitive.
Users with the correct key $k_e$ can regain the stega mask $\mathrm{M}_e$ locating the weights for secret data via our implicit consensus  (\cref{sub_method_1}). Thus, the secret data can be obtained through:
\begin{equation}
\mathbb{N}[\mathrm{W}\odot\mathrm{M}_e](\mathbf{u}) \to \hat{\mathbf{y}}_{se},
\end{equation}
where $\mathrm{W}\odot \mathrm{M}_e$ denotes the weights for secret representation. This $\mathrm{M}_e$ can be obtained with key $k_e$ via the \textbf{implicit consensus mechanism} strategy in \cref{sub_method_1}.

\begin{table*}[t]
\caption{Performance comparisons on different datasets. ``$\uparrow$": the larger the better, ``$\downarrow$": the smaller the better. Stega ratio $\mathcal{S}$ denotes the ratio of parameters used to represent secret data. The results of our U-INR are highlighted in cyan.}
\vspace{-0.3cm}
\setlength\tabcolsep{5.6pt}
\renewcommand{\arraystretch}{1.00} 
\centering
\resizebox{1.0\linewidth}{!}{
\begin{tabular}{@{}c|cccc|cccc|cccc@{}}
    \toprule
    \multirow{3}{*}{Methods}  & \multicolumn{12}{c}{Cover/Stega-image pair}  \\
    \cmidrule{2-13}
    &\multicolumn{4}{c|}{DIV2K}   &  \multicolumn{4}{c|}{COCO} & \multicolumn{4}{c}{ImageNet}  \\ 
    \cmidrule{2-13}
    &PSNR(dB)$\uparrow$&SSIM$\uparrow$&APD$\downarrow$&RMSE$\downarrow$ &PSNR(dB)$\uparrow$&SSIM$\uparrow$&APD$\downarrow$&RMSE$\downarrow$ &PSNR(dB)$\uparrow$&SSIM$\uparrow$&APD$\downarrow$&RMSE$\downarrow$ \\
    \cmidrule{1-13}
    HiDDeN~\cite{zhu2018hidden_r77}& 28.19& 0.9287& 8.01& 11.00& 29.16& 0.9318& 6.91& 9.60& 28.87& 0.9234& 7.43& 10.21 \\
    Baluja~\cite{Baluja2019image}& 28.42& 0.9347& 7.92& 10.64& 29.32& 0.9374& 7.04& 9.36& 28.82& 0.9303& 7.68& 10.21 \\
    HiNet~\cite{jing2021hinet}& 44.86& 0.9922& 1.00& 1.53& 46.47& 0.9925& 0.81& 1.30& 46.88& 0.9920& 0.81& 1.26 \\
    PUSNet~\cite{li2024purified_pusnet} &38.15& 0.9792& 2.30& 3.33& 39.09& 0.9772& 2.01& 2.96& 38.94& 0.9756& 2.21 & 3.06 \\
    \midrule
    \rowcolor{cyan!20}
    U-INR ($\mathcal{S} = 0.3$) &38.32 &0.9890 &2.17 &2.72  &39.70 &0.9889 &1.55 &2.39 &39.06 &0.9813 &1.61 &2.45 \\
    \rowcolor{cyan!20}
    U-INR ($\mathcal{S} = 0.5$) &35.15 &0.9740 &3.59 &4.70   &35.50 &0.9737 &3.55 &3.65 &35.32 &0.9770 &2.51 &3.63 \\
    \bottomrule
    \toprule
    \multirow{3}{*}{Methods} & \multicolumn{12}{c}{Secret/Recovered image pair} \\
    \cmidrule{2-13}
    &\multicolumn{4}{c|}{DIV2K}   &  \multicolumn{4}{c|}{COCO} & \multicolumn{4}{c}{ImageNet}  \\ 
    \cmidrule{2-13}
    &PSNR(dB)$\uparrow$&SSIM$\uparrow$&APD$\downarrow$&RMSE$\downarrow$ &PSNR(dB)$\uparrow$&SSIM$\uparrow$&APD$\downarrow$&RMSE$\downarrow$ &PSNR(dB)$\uparrow$&SSIM$\uparrow$&APD$\downarrow$&RMSE$\downarrow$ \\
    \midrule
    HiDDeN~\cite{zhu2018hidden_r77}& 28.42& 0.8695& 7.62& 9.94& 28.81& 0.8576& 7.20& 9.54& 28.23& 0.8435& 7.83& 10.47 \\
    Baluja~\cite{Baluja2019image}& 28.53& 0.9036 & 7.53& 10.66& 29.13& 0.9091& 6.61& 9.80& 27.63& 0.8909& 8.33 & 12.26\\
    HiNet~\cite{jing2021hinet}& 28.66& 0.8507& 7.25& 9.68& 28.08& 0.8181& 7.80& 10.49& 27.94& 0.8159& 8.03& 10.83 \\
    PUSNet~\cite{li2024purified_pusnet} & 26.88& 0.8363& 8.75& 11.95& 26.96& 0.8211& 8.71& 12.14& 26.28& 0.8028& 9.58& 13.43 \\
    \midrule
    \rowcolor{cyan!20}
    U-INR ($\mathcal{S} = 0.3$)  &34.50 &0.9727 &3.82 &5.26  &34.92 &0.9802 &3.53 &4.86 &34.03 &0.9701 &3.97 &5.53 \\
    \rowcolor{cyan!20}
    U-INR ($\mathcal{S} = 0.5$) &37.11 &0.9851 &2.84 &3.86 &37.53 &0.9873 &2.73 &3.41 &36.82 &0.9802 & 2.90 &4.12\\
    \bottomrule
\end{tabular}}
\vspace{-0.3cm}
\label{tab:comparison}
\end{table*}

\begin{figure*}[t]
\centering
\includegraphics[width=1.0\linewidth]{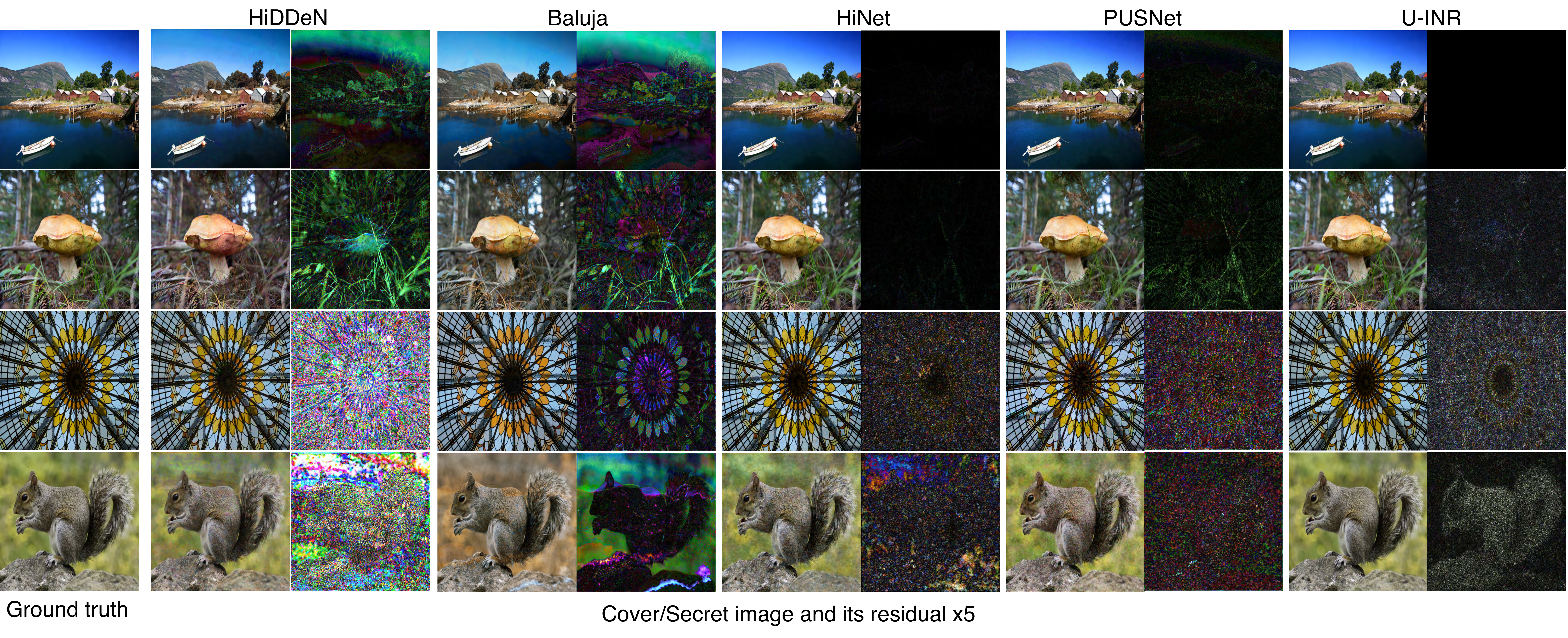}
\vspace{-0.85cm}
\caption{Examples of the stega and recovered images generated using different schemes. The left is the original image, and the right represents $\times 5$ magnified residuals. The cover/stega and secret/recovered images are given in the first and last rows. For our U-INR, we use stega ratio $\mathcal{S} = 0.3$ as it balances the quality of cover and secret representation.}
\vspace{-0.3cm}
\label{fig:image_comparision}
\end{figure*}
\section{Experiments}
\label{sec:exp}

\begin{figure*}[t]
\centering
\includegraphics[width=1.0\linewidth]{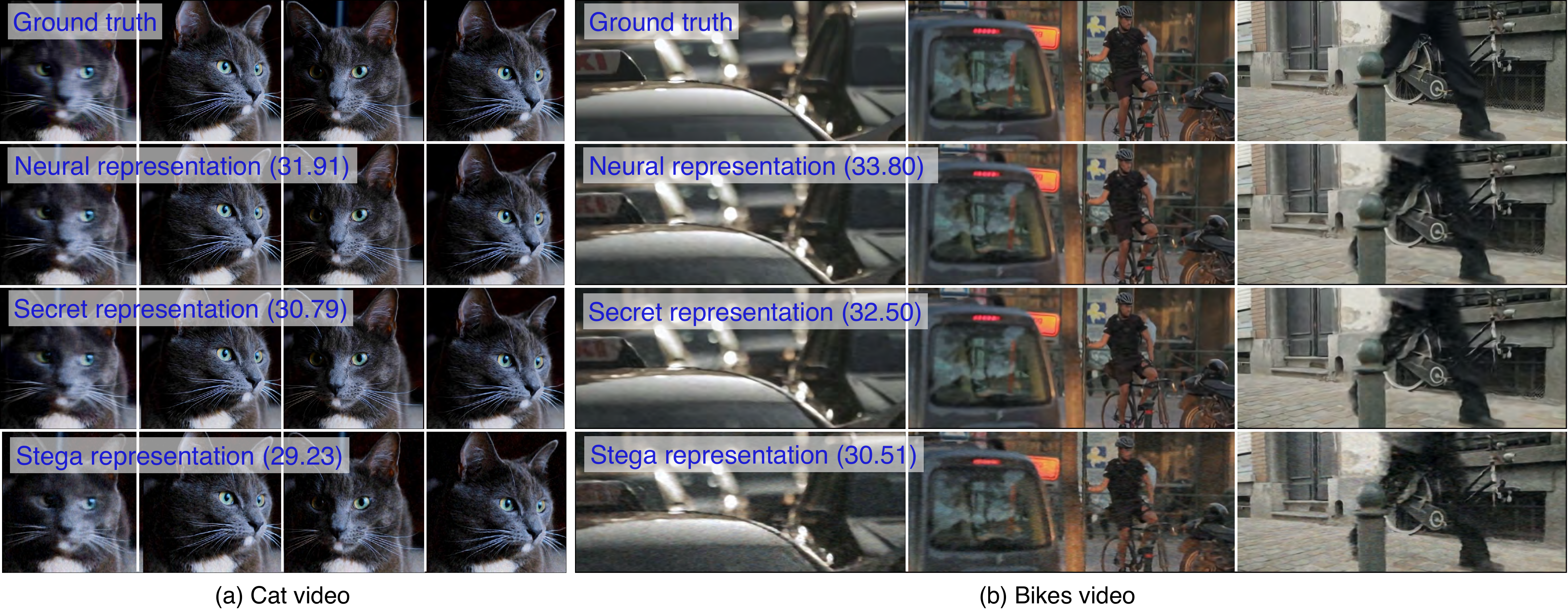}
\vspace{-0.8cm}
\caption{Quantitative and qualitative results of our method when applying to video data. The bike video and cat video are adopted as secret and stega representations. Compared to the normal neural representation, the quality of the stega and secret representations only experiences a slight decrease. }
\vspace{-0.0cm}
\label{fig:exp_vis}
\end{figure*}

\subsection{Experimental settings}

\subsubsection{Evaluation} To demonstrate the effectiveness of our method, we conduct experiments on different types of data, including images, videos, audio, and 3D scenes.
For images, following established work~\cite{li2024purified_pusnet}, we evaluate the performance on three dataset, including DIV2K~\cite{div2k}, $1,000$ images randomly selected from ImageNet~\cite{russakovsky2015imagenet}, and COCO~\cite{lin2014microsoft} dataset. For 3D scenes, we experiment with 4 classic scenes from LLFF~\cite{mildenhall2019llff} and NeRF-blender~\cite{mildenhall2020nerf}, respectively. We select the original test set in SIREN~\cite{sitzmann2020siren} for audio and video. We adopt PSNR, SSIM~\cite{wang2004image}, Averaged Pixel-wise Discrepancy (APD), and RMSE to measure the visual quality. Higher PSNR and SSIM values indicate better image embedding and recovery performance. Lower RMSE and APD values suggest improved performance in these tasks.  We adopt the MSE mean and MSE standard deviation for audio data to evaluate the quality of audio representation. We test the representation quality to evaluate the impact of the stega ratio $\mathcal{S}$ on the performance of the secret and stega representation under different ratios. Besides this, we also test the robustness of our method against network pruning~\cite{lee2021random_pruning,frankle2020pruning} to analyze potential threats.

\subsubsection{Benchmarks} To evaluate the performance of our method, we compare our U-INR against existing DNN-based steganographic methods, including HiDDeN~\cite{zhu2018hidden_r77}, Baluja~\cite{Baluja2019image}, HiNet~\cite{jing2021hinet}, and PUSNet~\cite{li2024purified_pusnet}. We implement the aforementioned models on the DIV2K training dataset for fair comparisons and evaluate their performance under the same settings~\cite{li2024purified_pusnet}. To illustrate the generalizability of our U-INR, we evaluate our U-INR on other types of data, including audio, video, 3D scene, and signed distance function.

\subsection{Results on different data types}

\subsubsection{Results on 2D image} To compare the capabilities of our method in steganography, we evaluate our U-INR and other image steganographic methods on various 2D image datasets~\cite{div2k,russakovsky2015imagenet,lin2014microsoft}. In \cref{tab:comparison}, we provide two sets of experimental results with different stega ratios $\mathcal{S} = \{0.3, 0.5\}$, denoting the ratios of the parameters for secret representation. Although these steganography methods~\cite{zhu2018hidden_r77,baluja2017hiding_rr5,jing2021hinet,li2024purified_pusnet} are specifically designed for 2D images, our U-INR achieves a better quality of the recovered secret image while keeping comparable results in the quality of the stega image.
Aligning with the visual samples in \cref{fig:image_comparision}, our U-INR achieves higher capacity steganography with a less obvious residual map. We successfully embed secret image representations into stega representations within the INR parameters while preserving the original image quality.

\begin{figure*}[ht]
\centering
\includegraphics[width=0.82\linewidth]{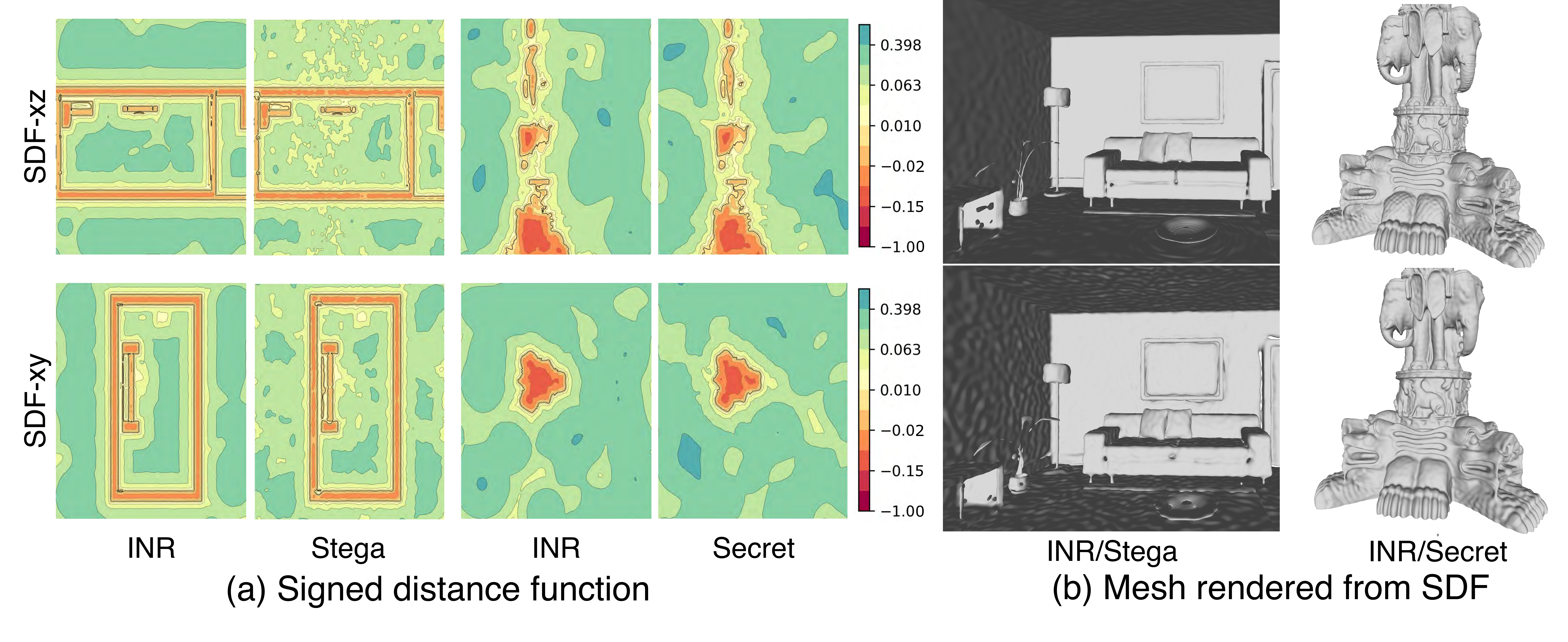} 
\vspace{-0.65cm}
\caption{A case study on Signed distance function. Thai statue (Secret) is extracted from the representation of Room (Stega). }
\vspace{-0.25cm}
\label{fig:exp_sdf}
\end{figure*}

\subsubsection{Results on video data} To demonstrate the generalizability of our approach, we extend our U-INR to video data and present the quantitative and qualitative results in \cref{fig:exp_vis}. We implement our U-INR on Cat and Bike videos from SiREN~\cite{sitzmann2020siren}. Comparing data as secret/stega representation with standard neural representation, our method minimally impacts quality. Our results showcase the ability to conceal and retrieve the hidden video content while maintaining the visual fidelity of the original video. Users can obtain hidden high-quality videos while ensuring invisibility. 

\subsubsection{Results on SDF data} We extend our U-INR to Signed Distance Function (SDF) data and present the results in \cref{fig:exp_sdf}. We apply our U-INR to the signed distance function and embed secret SDF representations within the network parameters. Our experimental results demonstrate the capability to conceal and retrieve the hidden SDF content while maintaining the geometric fidelity of the original structures. Users can obtain hidden high-quality SDF data while ensuring invisibility.

\begin{table}[t]
\caption{We implement our U-INR on 3D scenes with neural radiance field (NeRF)~\cite{mildenhall2020nerf}. We report the metric results when the scenes are adopted as stega representation and secret representation. The visualized results are presented in the appendix.
}
\vspace{-0.3cm}
\centering
\resizebox{0.85\linewidth}{!}{
\begin{tabular}{c|c|ccc}
    \toprule
& & PSNR(dB) $\uparrow$ &SSIM  $\uparrow$ &LPIPS $\downarrow$ \\
\midrule
&Lego &26.30 &0.9356 &0.1256\\
&Lego (Secret) &25.92 &0.9284 &0.1400 \\ 
\multirow{2.5}{*}{Blender}  &Lego (Stega) &25.62 &0.9182 &0.1511  \\
\cmidrule{2-5}
&Hotdog &32.86 &0.9778 &0.0671\\
&Hotdog (Secret) &32.45 &0.9751 &0.0782\\ 
&Hotdog (Stega) &32.40 &0.9733 &0.0830 \\
\midrule
&Flower &27.50 &0.8543 &0.1641\\
&Flower (Secret) &26.62 &0.8202 &0.2050\\ 
\multirow{2.5}{*}{LLFF} &Flower (Stega) &26.60 &0.8146 &0.2270 \\
\cmidrule{2-5}
&Room &30.64 &0.9313 &0.1674\\
&Room (Secret) &29.92 &0.9160 &0.2059\\ 
&Room (Stega) &29.74 &0.9109 &0.2265 \\
\bottomrule
\end{tabular}}
\vspace{-0.4cm}
\label{tab:3d_scene}
\end{table}

\subsubsection{Results on 3D scene} We implement our U-INR on 3D scene representations with Neural Radiance Fields (NeRFs)~\cite{mildenhall2020nerf}. We train the 3D presentation with multi-view images and embed secret 3D data within the parameters of NeRF models. Quantitative and qualitative results for each scene when used as stega and secret data are reported in \cref{tab:3d_scene} and \cref{fig:exp_vis_3d}. Compared to the original implicit neural representation, the quality of the stega and secret representations slightly declines, representing the 3D scene effectively. Our experiments demonstrate that our approach can effectively conceal and recover the hidden 3D content while preserving the high-quality rendering of the original 3D scenes. This capability opens up new avenues for secure data transmission and information hiding in the context of steganography 3D models.

\subsubsection{Results on audio data} To showcase the versatility of our approach, we extend our U-INR to audio data and present the results in Table \ref{tab:audio}. The visualized results are presented in the appendix.  We implement our U-INR on Bach and Counting audio~\cite{sitzmann2020siren}, embedding secret audio messages within the network parameters. The stega representations yielded enhanced performance, potentially attributable to the inherently lower complexity of audio data relative to other data modalities. Our results demonstrate the ability to conceal and retrieve the hidden audio content while preserving the auditory fidelity of the original signals. Users can obtain hidden high-quality audio while ensuring invisibility.

\subsection{Impacts of stega ratio} 
\begin{figure}[t]
\centering  
\includegraphics[width=0.90\linewidth]{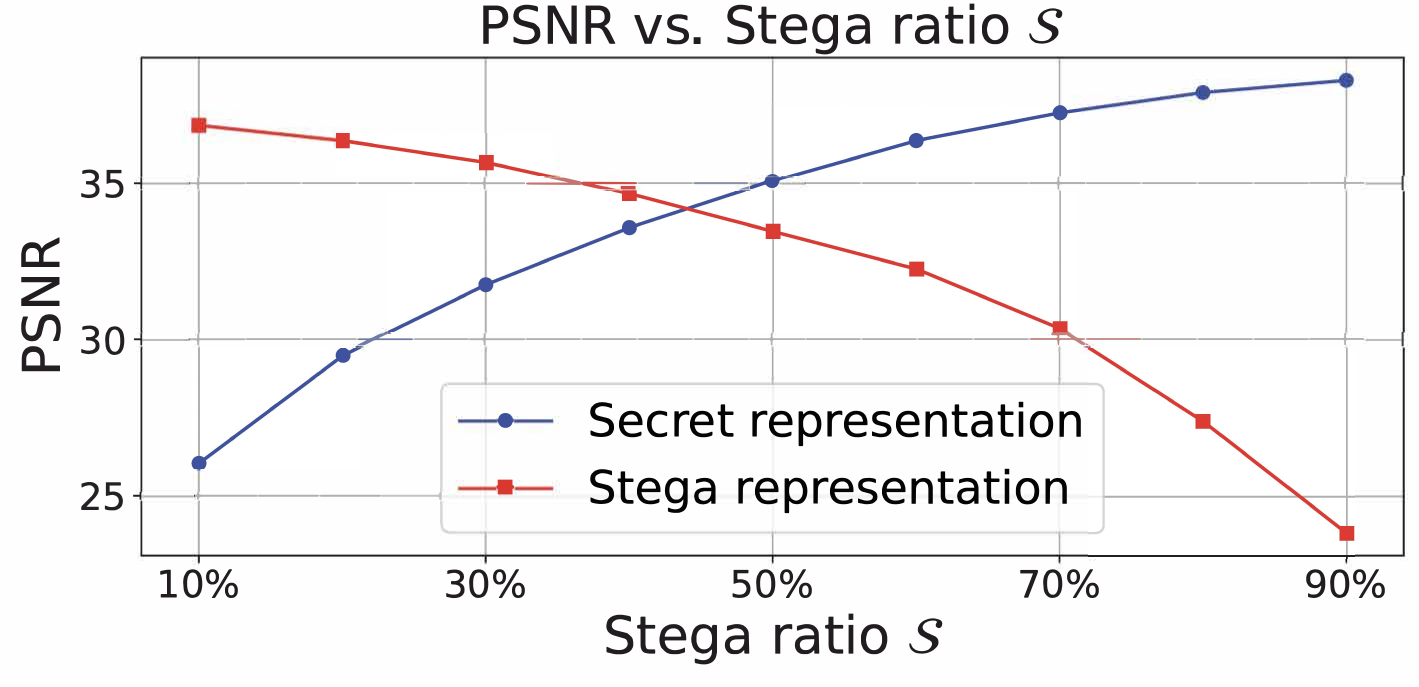} 
\vspace{-0.5cm}
\caption{Representation performance under different stega ratios. }
\label{fig:ratio_capability}
\vspace{-0.4cm}
\end{figure}

Stega ratio $\mathcal{S}$ indicates the percentage of INR parameters used for secret representation. Thus, selecting the right ratio depends on whether the priority is to preserve ``secret'' information or maintain broad coverage. We evaluate the impacts of selecting a suitable ratio for U-INR to identify the suitable ratio that trades off the secret and stega representation quality. We experiment across a range of sparsity levels from 10\% to 90\% on DIV2K test-set~\cite{div2k}. As shown in \cref{fig:ratio_capability}, a higher ratio will decrease stega quality and improve secret representation quality. The experimental results reveal a trade-off when adjusting the sparsity ratios, indicating the trade-off choice in the intermediate range of 30\% to 70\%. 

\begin{table}[t]
\caption{The audios Bach and Counting~\cite{sitzmann2020siren} are representation with INRs. The Bach and Counting are adopted as secret and stega representations, respectively. The experimental results are evaluated 10 times. }
\label{tab:audio}
\vspace{-0.2cm}
\centering
\resizebox{0.9\linewidth}{!}{
\begin{tabular}{ccccc}
    \toprule
& MSE Mean $\downarrow$ &MSE Standard Dev. $\downarrow$  \\
\midrule
Bach &$1.980 \times 10^{-4}$  & $5.527\times 10^{-4}$ \\
Bach (Secret) &$2.387\times 10^{-4}$ &$6.101 \times 10^{-4}$ \\ 
Bach (Stega)  &$2.236 \times 10^{-5}$ &$6.024\times 10^{-5}$ \\
\midrule
Counting &$8.834 \times 10^{-4}$ &$5.374 \times 10^{-3}$ \\ 
Counting (Secret) &$9.718\times 10^{-4}$ &$5.594\times 10^{-3}$ \\
Counting (Stega) &$5.728 \times 10^{-4}$ &$3.915 \times 10^{-3}$ \\
\bottomrule
\end{tabular}}
\vspace{-0.1cm}
\end{table}

\subsection{Threaten analysis}
Pruning techniques are commonly used to compress and speed up neural networks by removing redundant parameters, which may impact the integrity of secret data. To test the resilience of our U-INR method against potential threats, we simulate pruning attacks on INR networks with stega representation. We evaluate the robustness of the stega representation against two pruning strategies, including magnitude-based pruning~\cite{frankle2020pruning} and random pruning~\cite{lee2021random_pruning}. For instance, we progressively prune increasing percentages of the INR parameters for each method and measure the deterioration. As shown in \cref{tab:pruning}, with higher attack strength, the quality of the representation declines. However, the secret representation is more robust against both pruning strategies than the stega representation, indicating the resilience of the method against advanced adversarial attacks.

\begin{table}[t]
\caption{Results of pruning weights in INR. The secret* representation has not been erased by the magnitude-based pruning method, as its weights have larger weight values than the stega representation. We further presented the analysis in the appendix.}
\vspace{-0.35cm}
\centering
\resizebox{0.82\linewidth}{!}{
\begin{tabular}{cccccccc}
    \toprule
Pruning &  &\multirow{2}{*}{0\%} &\multirow{2}{*}{1\%} &\multirow{2}{*}{5\%} &\multirow{2}{*}{10\%} &\multirow{2}{*}{20\%}\\
strategy \\
\midrule
            &INR &38.48 &35.39 &30.61 &28.06 &22.06  \\
Random &Stega &33.46 &30.24 &25.83 &23.46 &17.85  \\
            &Secret &35.08 &32.31 &28.97 &26.86 &21.05 \\
\midrule
            &INR &38.48 &37.99 &35.42 &29.63 &22.45  \\
Magnitude &Stega &33.46  &33.43 &30.87 &25.07 &18.19  \\
            &Secret* &35.08 &33.86 &33.86 &33.86 &33.86 \\
    \bottomrule
\end{tabular}}
\label{tab:pruning}
\vspace{-0.3cm}
\end{table}

\section{Conclusion}
This paper presents U-INR, a novel unified steganographic method leveraging implicit neural representations (INRs) to embed covert data across diverse media types. Our approach uniquely incorporates secret information within a portion parameter of the INR network, facilitating secure and discreet data transmission. Comprehensive experiments across multiple data modalities demonstrate U-INR's exceptional capacity for embedding information securely while preserving INR fidelity. These results underscore the method's robustness and its potential for broad applications in enhancing data security and privacy.



\bibliographystyle{ACM-Reference-Format}
\bibliography{reference}


\section{Appendix}
\renewcommand{\thefigure}{A.\arabic{figure}}
\setcounter{figure}{0}

\subsection{Implementation details}
\label{sec:details}
In our experiment, we implement SIREN\footnote{\url{https://github.com/vsitzmann/siren}}~\cite{sitzmann2020siren} for image, video, audio, SDF data. We implement  NeRF\footnote{\url{https://github.com/yenchenlin/nerf-pytorch}}~\cite{mildenhall2020nerf} for 3D scenes. All experiments are conducted on Ubuntu $18.04$ with one NVIDIA V100. The weights $\mathrm{W}$ of INR are initialized using Xavier algorithm~\cite{glorot2010understanding}. Following the established works~\cite{sitzmann2020siren,li2024purified_pusnet}, we conduct our experiments using the same settings. For 2D images, the model processes 2D feature vectors $(x, y)$ with four hidden layers of $256$ units each, outputting a 3D RGB vector. It is trained on datasets like the DIV2K-test, a $1, 000$ subset of Imagenet-$1k$, and the COCO test set using the Adam optimizer with a batch size of $1$, a learning rate of \(1 \times 10^{-4}\), over $5,000$ epochs. For videos, it processes 3D feature vectors $(x, y, t)$ with three hidden layers of $1,024$ units each, outputting a 3D RGB vector trained on Cat and Bikes videos with similar optimization parameters over $100,000$ epochs. For audio, the model handles 1D feature vectors $t$ with three hidden layers of $256$ units each and an initial frequency \(\omega_0 = 3,000\), trained on Counting and Bach audio data~\cite{sitzmann2020siren} with a learning rate of \(1 \times 10^{-4}\) over $1,000$ steps. The SDF model processes 3D feature vectors $(x, y, z)$ through three hidden layers of 256 units, outputting a 1D distance value, trained on data from TurboSquid and the Stanford 3D Scanning Repository with a batch size of $1,400$ over $10,000$ steps. For 3D scenes, NeRF uses a 5D input vector $(x, y, z, \theta, \phi)$ with an 8-layer MLP of $256$ channels per layer, processing $4,096$ rays per batch, trained on LLFF and NeRF-blender datasets with a learning rate of \(5 \times 10^{-4}\) and a decay step of $500,000$, simplifying the original NeRF strategy.

\subsection{Additional experimental results}
\subsubsection{Impacts of stega ratios}
We also provide visualized results that showcase the ratio $\mathcal{S}$, a critical parameter. As shown in \cref{fig:sparse_ratio}, the sparsity patterns and their impact on the model's performance are depicted. The visuals aim to demonstrate how varying levels of stega ratio $\mathcal{S}$ affect the representation performance of INR, elucidating the role of the ratio.

\subsubsection{Impacts of pruning attack}
We also provide visualized results that showcase the impacts of the pruning ratio $\mathcal{S}$. As shown in \cref{fig:pruning_attack}. The results demonstrate how varying levels of pruning ratio $\mathcal{S}$ affect the representation performance of stega and secret representation.

\subsubsection{Visualized audio data} \cref{fig:app_audio} presents the visualization results of the audio data to complement the study. These visualizations illustrate the characteristics and patterns within the audio samples. Representations of audio data are provided in \cref{fig:app_audio}, serving as a visual reference to understand our method for the audio data better.

\begin{figure}[h]
\centering
\includegraphics[width=0.80\linewidth]{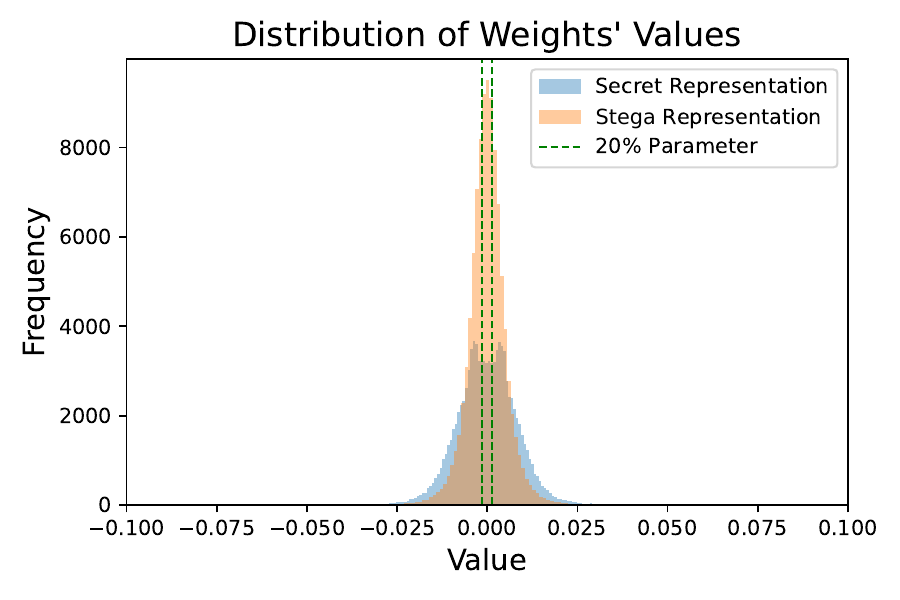} 
\caption{Distribution of weight's values. The weights for the stega representation are relatively small. Thus, during magnitude-based pruning, there is a tendency to prune the weights associated with the stega representation more. This will have a greater impact on the stega representation and a limited effect on the hidden secret representation.}
\label{fig:weights_dis}
\end{figure}

\subsection{Distribution of weight's values} To better understand our method, we have provided the distribution of weight values for ``stega'' and ``secret representation''. A standard SIREN~\cite{sitzmann2020siren} network for image representation has approximately $190, 000$ parameters. We visualize the range of parameter aggregation from $-0.1$ to $0.1$ and provide the distribution of weight's values in \cref{fig:weights_dis}. We observe that within the stega representation, the weight values of the secret representation are relatively high. Therefore, in magnitude-based pruning, the stega representation is more significantly affected. We present a series of results to portray the effects of pruning attacks on the system \cref{fig:pruning_attack} and illustrate the outcomes of these attacks, emphasizing the resilience of the proposed method. The visuals document the robustness of the approach, providing a breakdown of the system's capability to withstand pruning attacks and maintain performance integrity.

\begin{figure*}[h]
\centering
\includegraphics[width=1.0\linewidth]{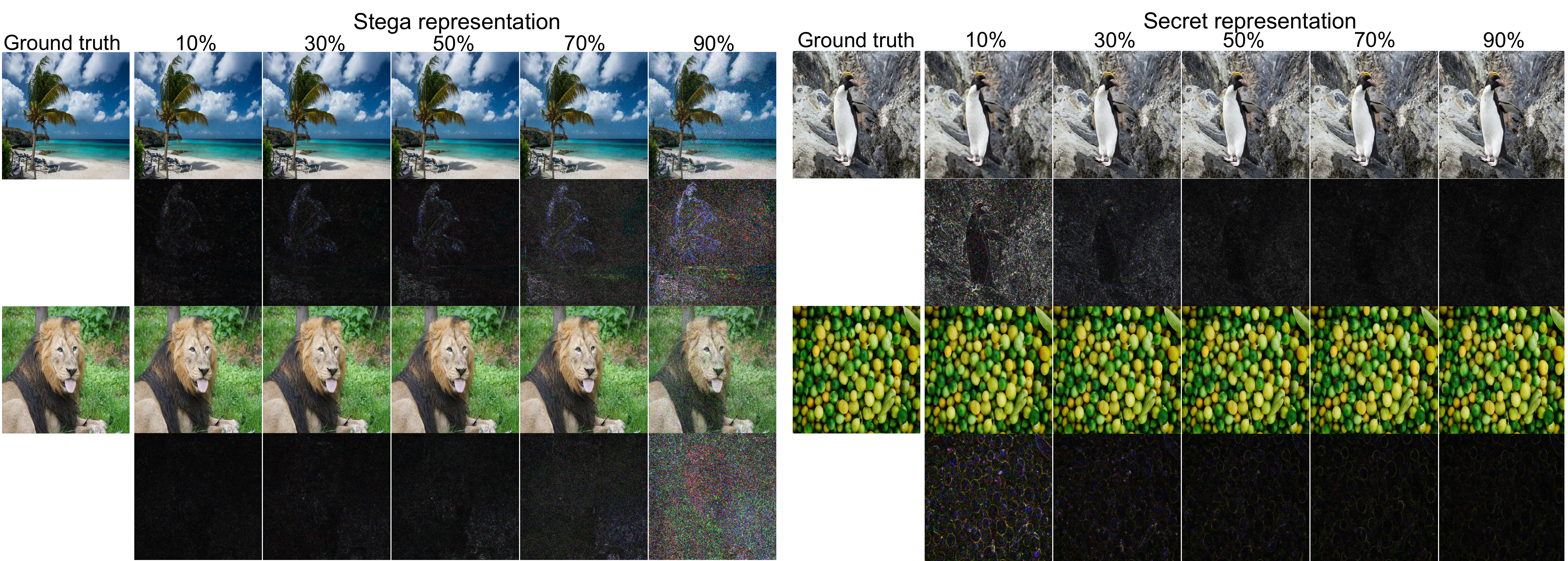} 
\caption{Qualitative results of \textbf{stega representation} and \textbf{secret representation} with different stega ratios $\mathcal{S}$. $\mathcal{S}$ denotes the ratio of parameters used for representing secret data. The residual images ($\times 5$) are located beneath each picture.}
\label{fig:sparse_ratio}
\end{figure*}

\begin{figure*}[h]
\centering
\includegraphics[width=0.95\linewidth]{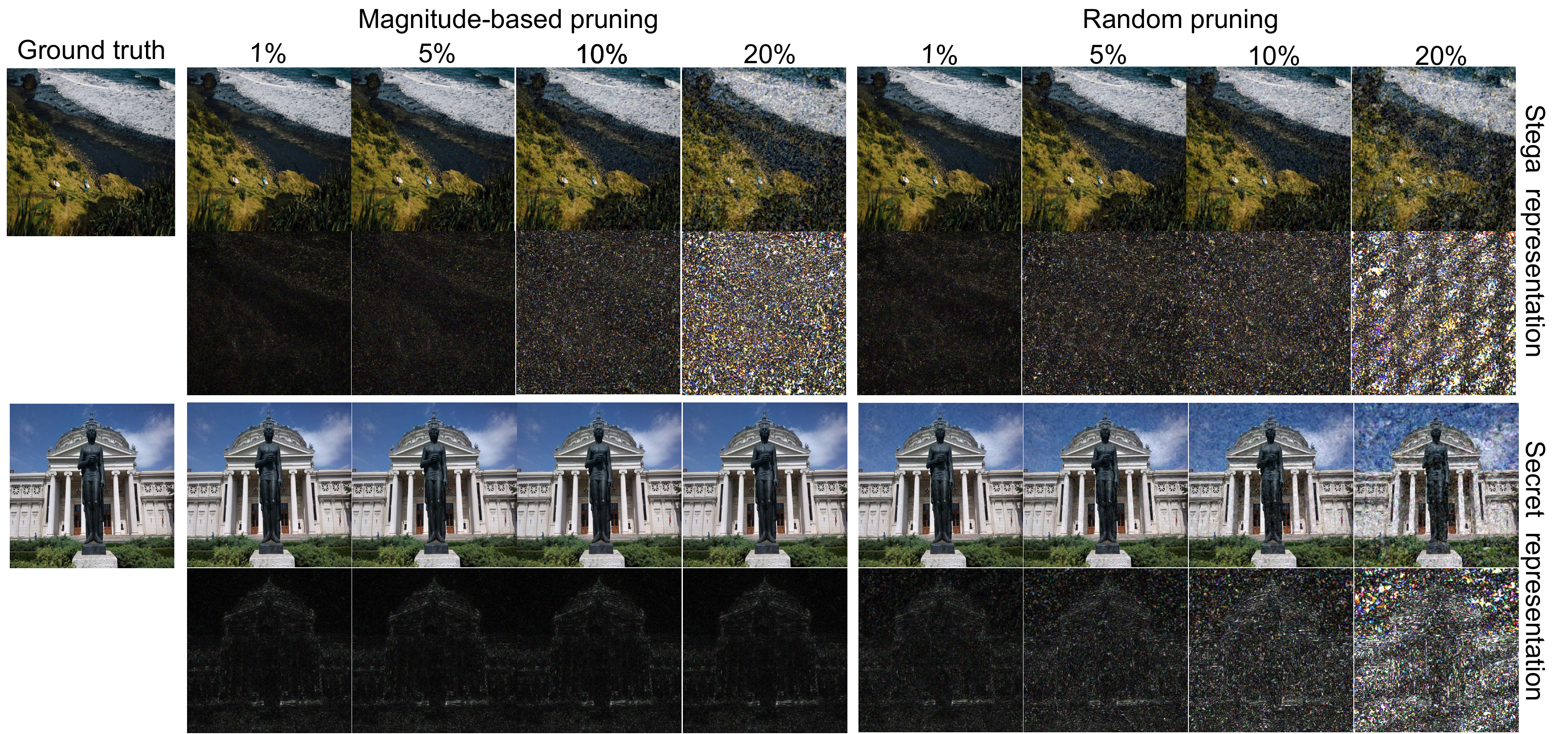} 
\caption{Qualitative results of \textbf{random pruning} and \textbf{magnitude-based pruning}  attack on INR, stega, and secret representation. Each neural network for representation is randomly pruning 1\%, 5\%, 10\%, and 20\% parameters. The residual images ($\times 5$) are located beneath each picture.}
\label{fig:pruning_attack}
\end{figure*}

\begin{figure*}[h]
\centering
\includegraphics[width=1.0\linewidth]{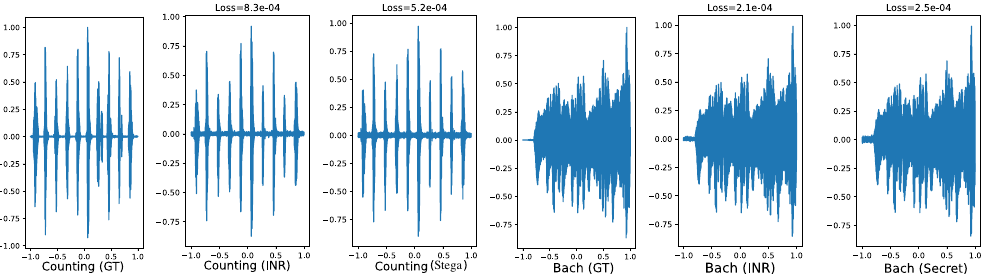} 
\caption{A case study on audio format data. Bach (secret) is obtained from the representation of Counting (stega).  The audio Counting (stega) is better than Counting (INR), which might be due to the audio data being simpler than other data formats.}
\label{fig:app_audio}
\end{figure*}

\begin{figure*}[h]
\centering
\includegraphics[width=0.98\linewidth]{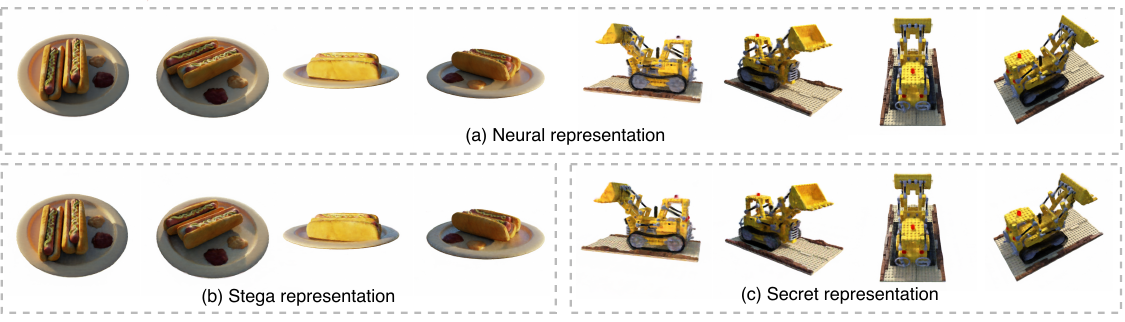} 
\caption{Visual results of our proposed method on 3D scene~\cite{mildenhall2020nerf}. We embed a secret representation (lego) into the stega representation (hotdog). (a) INR-based representation denotes the baseline results of NeRF when rendering both scenes. Scene (c) secret representation is obtained from (b) stega representation.}
\label{fig:exp_vis_3d}
\end{figure*}

\end{document}